\documentclass[aip,jcp,preprint]{revtex4-1}
\usepackage{graphicx}
\usepackage{amsfonts}
\usepackage{amsmath}
\usepackage{bm}
\usepackage[usenames,dvipsnames]{color}

\begin{document}
\title{Rotational predissociation of extremely weakly bound atom-molecule complexes produced by Feshbach resonance association}
\author{Alisdair O. G. Wallis}
\affiliation{Department of Chemistry, University of British Columbia,
Vancouver, British Columbia, V6T~1Z1, Canada}
\author{Roman V. Krems}
\affiliation{Department of Chemistry, University of British Columbia,
Vancouver, British Columbia, V6T~1Z1, Canada}
\date{\today}
\begin{abstract}
We study the rotational predissociation of atom - molecule complexes with very small binding energy. Such complexes can be produced by Feshbach resonance association of ultracold molecules with ultracold atoms. Numerical calculations of the predissociation lifetimes based on the computation of the energy dependence of the scattering matrix elements become inaccurate when the binding energy is smaller than the energy width of the predissociating state. 
We derive expressions that represent accurately the predissociation lifetimes in terms of the real and imaginary parts of the scattering length and effective range for molecules in an excited rotational state. Our results show that the predissociation lifetimes are the longest when the binding energy is positive, i.e. when the predissociating state is just above the excited state threshold. 

\end{abstract}
\maketitle




\section{Introduction}

The energy levels of ultracold atoms can be shifted by an external magnetic field. This gives rise to shifts of bound energy levels of atom - atom complexes. Under certain conditions, an adiabatic variation of the external field can be used to move a scattering state of ultracold atoms to below threshold. This results in the formation of atom - atom complexes with extremely small binding energy ($< 10^{-8}$ eV). This technique -- known as Feshbach-resonance association -- is widely used for the production of ultracold molecules from ultracold atoms  \cite{Doyle:2004,Hutson:IRPC:2006,Kohler:RMP:2006,Carr:NJPintro:2009,Chin:RMP:2010}. Feshbach resonance association can also be used to link ultracold atoms with ultracold molecules or ultracold molecules with ultracold molecules \cite{Chin:2005} to produce weakly bound atom - molecule or molecule - molecule complexes. The binding energy of such complexes can be tuned from zero to about $0.5$ cm$^{-1}$ by varying an external field. If the molecules are initially in a ro-vibrationally excited state, these weakly bound complexes may undergo ro-vibrational predissociation. While the ro-vibrational predissociation of van der Waals atom - molecule complexes has been the subject of numerous studies \cite{Beswick:1978,Pine:1983,Ashton:1983,giancarlo:1994}, little is known about the predissociation of atom - molecule complexes with binding energies $< 0.5$ cm$^{-1}$. In the present paper, we explore the dependence of the rotational predissociation lifetimes of weakly bound atom - molecule complexes on the binding energy in the limit of zero binding energy.

The present work is motivated by two recent research developments. First, the creation of ultracold molecules in the ro-vibrational ground state, achieved by several research groups \cite{Lang:ground:2008,Ni:KRb:2008},  makes the measurements of near-threshold rotational predissociation possible. The predictions of the present work can therefore be readily tested in experiments with ultracold molecules. Second, several recent studies showed that an ensemble of ultracold molecules trapped on an optical lattice may undergo collective rotational excitations analogous to Frenkel excitons in solid-state molecular crystals \cite{Herrera:2010,  Perez:2010, Herrera:excitionphonon:2010}. Ultracold molecules on an optical lattice can therefore be used as a model for quantum simulation of energy transfer in solid-state crystals and molecular aggregates \cite{Herrera:excitionphonon:2010}. Of particular interest would be a study of energy transfer in the presence of a disorder potential \cite{Herrera:2010}. A disorder potential can be introduced 
to an ensemble of molecules on an optical lattice by adding atoms to certain lattice sites. If the atoms are coupled to molecules by Feshbach resonance association, the atom - molecule interactions may introduce controllable perturbations affecting the Frenkel excitons. The rotational predissociation of atom - molecule complexes would, however, lead to loss of molecules from the optical lattice, thus limiting the experiments. It is therefore important to explore if the lifetimes for the rotational predissociation can exceed the timescales for the rotational energy transfer between molecules on an optical lattice \cite{Perez:2010}.

The dependence of the predissociation lifetimes on the binding energy can be qualitatively understood using perturbation theory \cite{LeRoy:1982,HUTSON:ArH2:1983,Forrey:vib:1999}. Using the unperturbed single channel radial functions for the bound and continuum states, it can be shown that the predissociation lifetime is inversely proportional to the square root of the binding energy.  However, perturbative calculations fail to give quantitative results due to the difficulty of representing the near-threshold bound state wavefunctions and the neglect of couplings between continuum states. An alternative approach is based on multichannel effective range theory and the calculation of elastic and inelastic scattering matrix elements for the excited state in the limit of zero collision energy \cite{Balakrishnan:scat-len:1997,Forrey:1998}.  This method can be used to estimate the location and lifetime of the most weakly bound predissociating state.  However, all previous calculations were based on the assumption that the couplings between molecular states are exceedingly small, resulting in exceedingly weak inelastic scattering.  In the vicinity of a Feshbach resonance, giving rise to a predissociating state near threshold, this assumption is generally not valid. In this paper, we extend the work of Balakrishnan and coworkers \cite{Balakrishnan:scat-len:1997} and Forrey and coworkers \cite{Forrey:1998} to derive the relations between the predissociation lifetime and the scattering parameters of the atom - molecule complex without any assumptions about the strength of inelastic scattering.  We calculate the predissociation lifetimes for the prototypical Mg($^1S$)+NH($^3\Sigma^-,n=1$) ultracold system, where $n$ is the molecular rotational angular momentum. We find that near threshold the predissociation lifetime can be as large as 100 $\mu$s.

\section{Theory}


At low collision energies, the scattering cross section of an atom and a molecule is entirely determined by the $s$-wave contribution to the partial wave expansion and the scattering observables can be expressed in terms of a few parameters. For simplicity of discussion, we assume that the atom is structureless. Using effective-range theory \cite{Mott:1965}, the first few terms in a power series expansion of the cotangent of the scattering phase shift $\delta_i$, or equivalently the diagonal $S$-matrix element $S_{ii}=\exp(2i\delta_i)$, for a molecular state $i$, can be written as
\begin{equation}\label{eqn:cotd}
k_i\cot\delta_i =-ik_i\left(\frac{1+S_{ii}}{1-S_{ii}}\right)
= -\frac{1}{a_i}+\frac{r_{i} k_i^2}{2}+\mathcal{O}(k^4),
\end{equation}
where $k_i$ is the scattering channel wave number defined by $\hbar^2k_i^2=2\mu (E-E_i)$, $\mu$ is the reduced mass, $E$ is the total energy and $E_i$ is the energy of the molecular state $i$. Each molecular state gives rise to a scattering channel. The constants $a_i$ and $r_i$ are the channel scattering length and effective range, respectively. The atom-molecule interaction potential depends asymptotically on the atom - molecule separation $R$ as $R^{-n}$. The scattering length exists if $n>3$ and the effective range exists if $n>5$.

If state $i$ is the absolute ground state of the molecule, only elastic scattering is possible, $|S_{ii}|=1$ and both the scattering length and the effective range are real. 
If state $i$ is an excited state, inelastic scattering becomes possible, $|S_{ii}|\le1$ and both the scattering length and the effective range become complex
\cite{Ross:1960,Ross:1961,Balakrishnan:scat-len:1997,Bohn:1997}, 
$a_i=\alpha_i-i\beta_i$ and $r_i = r_{\text{R},i}+ir_{\text{I},i}$.
Indeed for two $s$-wave states, a ground and an excited state, the energy dependence of the scattering observables near the excited state threshold can be characterized by just six real parameters: the real and imaginary parts of the scattering length and effective range of the excited state, the tangent of the ground state scattering phase shift at threshold and an additional effective range parameter that describes the first order energy dependence of the ground state phase shift at threshold.


A Feshbach resonance occurs when a bound state corresponding to an asymptotically closed channel ($E_i>E$) is the same as the collision energy of lower-energy molecular states.  
In the single channel case, as the system is tuned across a resonance, the $S$-matrix element describes a unit circle in the complex plane and the scattering length exhibits a pole when $S_{ii}$ passes through $-1$. In the multichannel case, $S_{ii}$ no longer describes a unit circle in the complex plane and thus the scattering length no longer exhibits a pole but an oscillation \cite{Hutson:res:nonote:2007}.
The location of the Feshbach resonance corresponds to a pole in $S_{ii}$.  Regarding $S_{ii}$ as a function of the complex energy $\mathcal{E}$ \cite{Taylor:1972}, the pole occurs at
\begin{equation}
\mathcal{E}=E_{\rm B}-i\frac{\Gamma}{2},
\end{equation}
where $E_\text{B}$ is the location of the bound state/resonance and $\Gamma$ is the resonance width, related to the bound state lifetime by $\tau=\hbar/\Gamma$.

To allow analysis below threshold (at negative channel energies), we define the complex channel wave number $\kappa_i=-ik_i$ such that
\begin{equation}
\mathcal{E}=-\frac{\hbar^2\kappa_i^2}{2\mu}.
\end{equation}
Balakrishnan et al. \cite{Balakrishnan:scat-len:1997} related the location and lifetime of the bound state/resonance below threshold to the complex scattering length calculated above threshold. In subsequent work \cite{Forrey:1998}, the treatment was expanded to include the complex effective range. However in this latter work only the limiting case of small inelasticity was considered, in which $\alpha\gg\beta$ and $r_{\text{R},i}\gg r_{\text{I},i}$. In the vicinity of a Feshbach resonance, the pole in the real part of the scattering length is suppressed \cite{Gonzalez-Martinez:2007,Hutson:res:nonote:2007} and thus the magnitude of $\beta$ may be comparable with or greater than that of $\alpha$ (see Figure 1, for example).

The location of a Feshbach resonance also corresponds to a pole in the scattering amplitude $f(k)$. Following Balakrishnan et al. \cite{Balakrishnan:scat-len:1997}, the scattering amplitude can be written as
\begin{equation}
f(k)=\left[-\frac{1}{a_i}-\frac{r_i}{2}\kappa_i^2+\kappa_i\right]^{-1},
\end{equation}
for a potential that asymptotically decays faster than $R^{-4}$. In the limit of $\kappa_i\rightarrow0$, the pole occurs when $\kappa_i=1/a_i$ and thus when
\begin{equation}\label{eqn:a}
\mathcal{E}=-\frac{\hbar^2}{2\mu a_i^2},
\end{equation}
and the bound state energy and width are explicitly
\begin{eqnarray}\label{eqn:scatlen}
E_{\rm B} = -\frac{\hbar^2}{2\mu |a|^4}\left(\alpha^2-\beta^2\right) \\
 \Gamma = \frac{\hbar}{\tau}=\frac{2\hbar^2\alpha\beta}{\mu|a|^4} .
\end{eqnarray}
%
In the single-channel case ($\beta=0$), 
as the bound state goes from just below to just above threshold, at which point the bound state becomes virtual, the scattering length goes through a pole changing sign from $+\infty$ to $-\infty$ \cite{Moerdijk:1995}. In the presence of inelastic channels ($\beta\neq0$) the bound state is coupled to the other channels which leads to a shift in the bound state energy. As the bound state is tuned from just below threshold to just above, the bound state crosses threshold ($E_\text{B}=0$) when $\alpha^2=\beta^2$ and becomes a virtual state ($\Gamma<0$) when $\alpha<0$ (or $\beta<0$). 
In the presence of inelastic channels the true bound state can thus exist at energies greater than the asymptotic threshold energy. 
%

Incorporating the effective range as a correction to (\ref{eqn:a}) \cite{Forrey:1998}, the pole in $f(k)$ occurs when
\begin{equation}
\kappa_i=\frac{1}{r_i}-\sqrt{\frac{1}{r_i^2}-\frac{2}{a_ir_i}}.
\end{equation}
Making no assumptions about the nature of the scattering, the complex energy at which the pole occurs is
\begin{equation}\label{eqn:aandr}
\mathcal{E}=-\frac{\hbar^2}{\mu r_i^2}\left(1-\frac{r_i}{a_i}-\sqrt{1-\frac{2r_i}{a_i}}\right).
\end{equation}
%
Introducing the notation
\begin{equation}
\rho+i\sigma=(r_{\rm R}+ir_{\rm I})(\alpha+i\beta)
=\alpha r_\text{R}-\beta r_\text{I} +i(\alpha r_\text{I}+\beta r_\text{R})
\end{equation}
and 
\begin{eqnarray}
\sqrt{1-\frac{2(\rho+i\sigma)}{|a|^2}}=
\left(\sqrt{\frac{1}{4}-\frac{\rho}{|a|^2}+\frac{\rho^2+\sigma^2}{|a|^4}}+\frac{1}{2}-\frac{\rho}{|a|^2}\right)^\frac{1}{2}  \nonumber \\
-i\left(\sqrt{\frac{1}{4}-\frac{\rho}{|a|^2}+\frac{\rho^2+\sigma^2}{|a|^4}}-\frac{1}{2}+\frac{\rho}{|a|^2}
\right)^\frac{1}{2}
=S_\text{R}-iS_\text{I},
\end{eqnarray}
the complex energy at which the pole occurs can be explicitly written as
\begin{equation}
\mathcal{E}=-\frac{\hbar^2}{\mu}\frac{ (r_\text{R}^2-r_\text{I}^2)-2ir_\text{R}r_\text{I}}{|r_i|^4}
\left( 1 -\frac{\rho+i\sigma}{|a|^2}-S_\text{R}+iS_\text{I}\right),
\end{equation}
where the location and width of the bound state are
\begin{eqnarray}\label{eqn:effrange}
E_\text{B} = -\frac{\hbar^2}{\mu}\left\{
\frac{(r_\text{R}^2-r_\text{I}^2)}{|r_i|^4}\left(1-\frac{\rho}{|a|^2}-S_\text{R}\right)
-\frac{2r_\text{R}r_\text{I}}{|r_i|^4}\left(\frac{\sigma}{|a|^2}-S_\text{I}\right)
\right\}
 \\
\Gamma =\frac{2\hbar^2}{\mu}\left\{
\frac{(r_\text{R}^2-r_\text{I}^2)}{|r_i|^4}\left(S_\text{I}-\frac{\sigma}{|a|^2}\right)
-\frac{2r_\text{R}r_\text{I}}{|r_i|^4}\left(1-\frac{\rho}{|a|^2}-S_\text{R}\right)
\right\}.
\end{eqnarray}

\section{Numerical Calculations}


To test the validity of equations (\ref{eqn:a}) and (\ref{eqn:aandr}), 
we numerically calculate the location and width of the
highest energy $s$-wave bound states of a Mg($^1S$) atom bound to a rotationally excited NH($^3\Sigma^-$) molecule.
The potential energy surface of the Mg+NH complex was calculated by Sold\'an, \.Zuchowski and Hutson \cite{Soldan:MgNH:2009}. This potential energy surface was used in quantum mechanical scattering calculations  \cite{Wallis:PRL:MgNH:2009} demonstrating that the sympathetic cooling of NH by Mg has a good prospect of success. The Mg-NH energy surface is anisotropic, providing significant couplings between the predissociating and continuum states. 

The rotational structure of NH exhibits fine structure due to the coupling of the molecular spin ${\bm s}$ to the rotational angular momentum ${\bm n}$. The molecular energy levels are characterized by the quantum number of the total angular momentum ${\bm j}={\bm n}+ {\bm s}$ of the molecule. There is thus a single ground rotational state $|n=0,j=1\rangle$ and three states $|n=1,j=0,1,2\rangle$ corresponding to $n=1$.
%
The end-over-end rotation of the complex is described by the angular momentum ${\bm L}$ which couples to ${\bm j}$ to form the total angular momentum ${\bm J}={\bm j}+{\bm L}$. The matrix of the total Hamiltonian describing the Mg - NH complex can be represented in the fully coupled angular momentum basis set $|(ns)jLJM\rangle$, as described in \cite{Wallis:phd:2010}. We perform the scattering calculations using the MOLSCAT package \cite{molscat:v14}, in which sets of coupled equations are constructed for each $J$ and parity $(-1)^{n+L+1}$ in the fully coupled basis set $|(ns)jLJM\rangle$ with $n\le6$ and $L\le12$. The coupled differential equations are integrated using the hybrid log-derivative method of Alexander and Manolopoulos \cite{Alexander:1987} based on a fixed-step-size log-derivative propagator in the short-range region (2.5 to 50 \AA) and a variable-step-size Airy propagator in the long-range region (50 to 250 \AA). The log-derivative solution is matched to the asymptotic boundary conditions \cite{Johnson:1973} to obtain the $S$-matrix.
%
To calculate the scattering length and effective range for use in equations (\ref{eqn:a}) and (\ref{eqn:aandr}), the $S$-matrix is calculated above threshold over the energy range 10 pK to 100 nK  and then fitted to equation (\ref{eqn:cotd}).

\begin{figure}
\begin{center}
\includegraphics[width=0.95\linewidth]{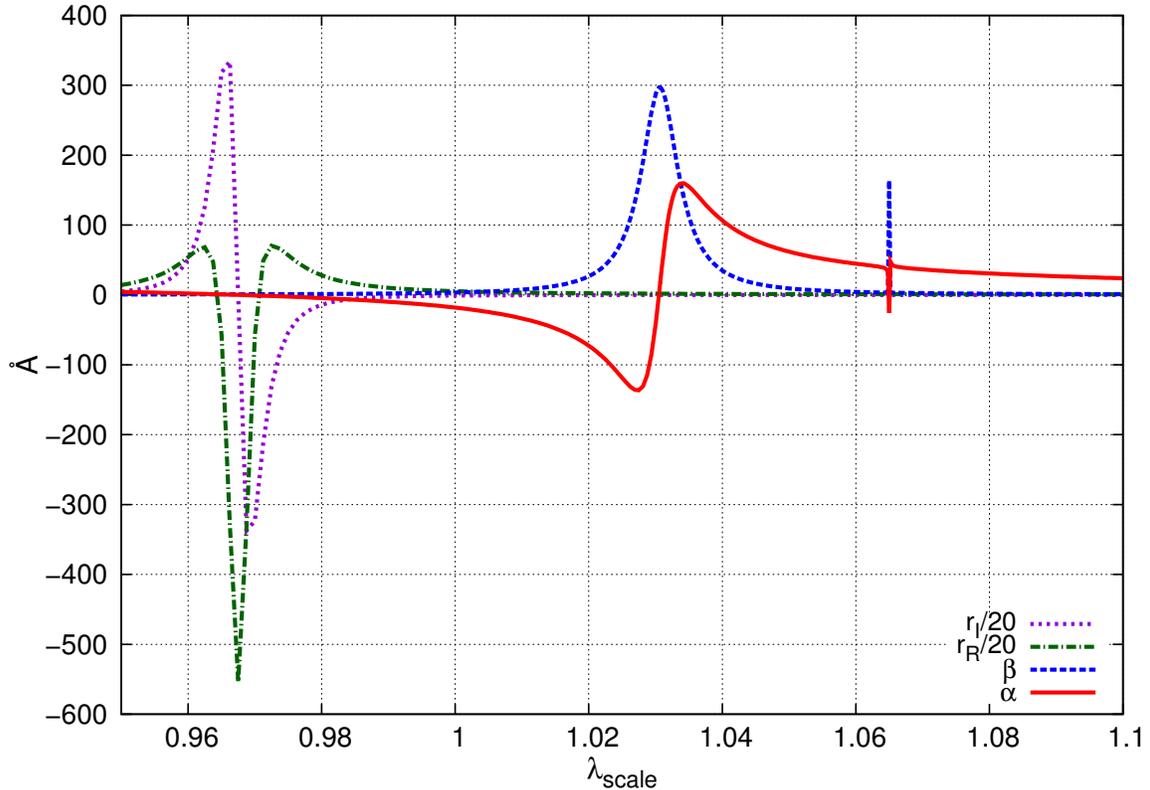}
\caption{The real and imaginary parts of 
the Mg+NH($n=1,j=1$) $s$-wave scattering length $a=\alpha-i\beta$ and effective range $r=r_\text{R}+ir_\text{I}$ as functions of the 
scaled interaction potential, are shown as $\alpha$ (red, solid line), $\beta$ (blue, dashed line), $r_\text{R}$ (green, dashed-dotted line), and $r_\text{I}$ (purple, dotted line).}
\label{fig:potdep}
\end{center}
\end{figure}

To vary the location of bound states with respect to threshold, we introduce a scaling factor to the potential
\begin{equation}
V^{\rm scaled}(\mathbf{R})=\lambda_\text{scale}V(\mathbf{R}).
\end{equation}
In the following, we consider the first (most weakly bound) predissociating state below the Mg+NH$(n=1,j=1),L=0$ threshold. Figure~\ref{fig:potdep} shows the complex scattering length and effective range as a function of $\lambda_\text{scale}$. As the bound state energy crosses threshold (at $\lambda_\text{scale}\approx1.03$), the real and imaginary parts of the scattering length show the expected oscillation and peak respectively. The real part of the effective range also shows a suppressed pole as $\alpha$ passes non-resonantly through zero.
An additional resonance is also present at  $\lambda_\text{scale}\approx1.065$, due to an $n=2$ state crossing threshold.

\begin{figure}
\begin{center}
\includegraphics[width=0.47\linewidth]{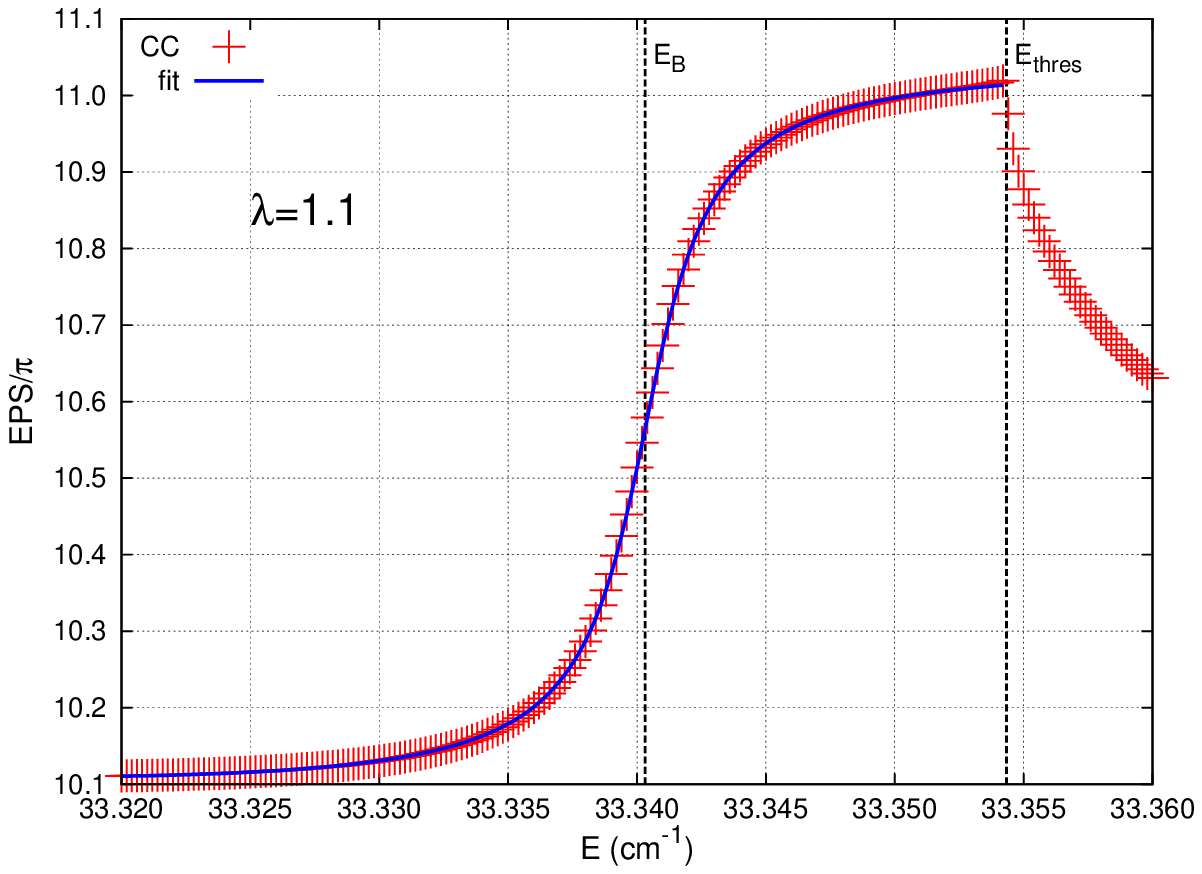}
\includegraphics[width=0.47\linewidth]{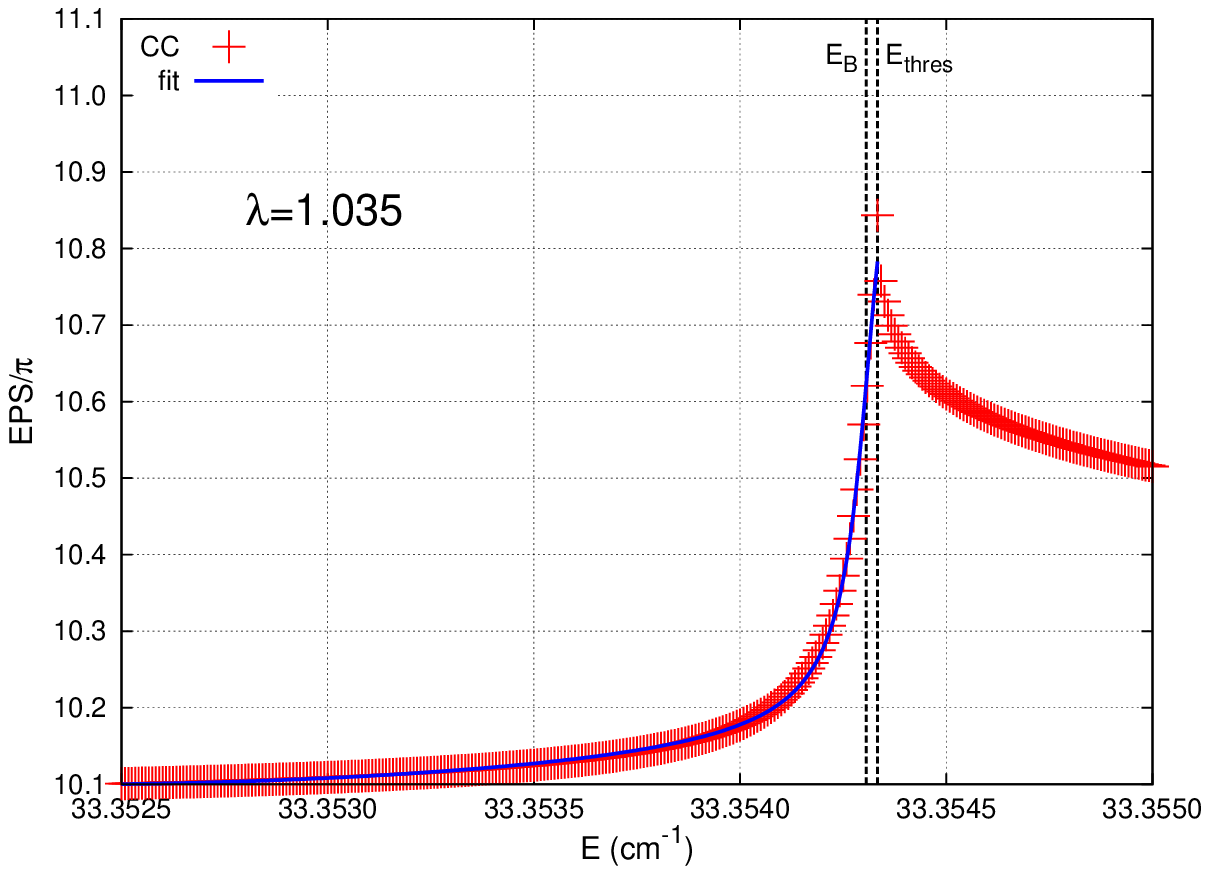}
\caption{Left panel: The location and width of a predissociating bound state at energy $E_\text{B}$ well below the threshold energy $E_{\rm thres}$ 
 is calculated by fitting the  Breit-Wigner form (blue, solid line) to the $S$-matrix eigenphase sum (EPS), obtained from solving a set of coupled-channel (CC) equations (red, crosses). Right panel: For a predissociating state close to ($\Gamma>E_\text{B}-E_\text{thres}$) or above threshold, as shown for $\lambda_\text{scale}=1.035$, the location and width of the state are extrapolated from the eigenphase sum below threshold.}
\label{fig:fitting}
\end{center}
\end{figure}

The location and width of each predissociating bound state 
are determined by calculating the $S$-matrix as a function of energy. 
 For a single open channel, the scattering phase shift $\delta_i$ goes through $\pi$ at the bound state energy following the Breit-Wigner energy dependence
\begin{equation}
\delta_i(E)=\delta_{\rm bg}(E)+\tan^{-1}\left(\frac{\Gamma}{2(E_{\rm B}-E)}\right),
\end{equation}
where $\delta_{\rm bg}(E)$ is a slowly varying background term. In the multichannel case, 
the single channel phase shift no longer follows to Breit-Wigner form and 
it is the $S$-matrix eigenphase sum $\sigma(E)$ \cite{Ashton:1983}, the sum of the phases of the eigenvalues of the  $S$-matrix, that follows the Breit-Wigner form. 
%
When the bound state exists well below threshold, $\Gamma < E_{\rm thres}-E_\text{B}$, the width and location are found by fitting the eigenphase sum across resonance to the Breit-Wigner equation using the RESFIT computer program \cite{Hutson:resfit:2007}, as shown for $\lambda_\text{scale}=1.1$ in figure~\ref{fig:fitting}. 
%
%
When the bound state exists close to threshold $\Gamma > | E_{\rm thres}-E_\text{B} |$ or just above threshold $E_\text{B}>E_{\rm thres}$, the threshold cusp prevents the eigenphase sum from going through the full Breit-Wigner form. An example is shown in figure~\ref{fig:fitting} for $\lambda_\text{scale}=1.035$ at which point the predissociating state is just below threshold.
The location and width of the resonance must thus be extrapolated from the eigenphase sum below threshold. Although the calculated eigenphase sum values are essentially exact, this limits the accuracy of the locations and widths obtained from the close coupling calculations for bound states very near (below or above) threshold. 

\begin{figure}[t]
\begin{center}
\includegraphics[width=0.9\linewidth]{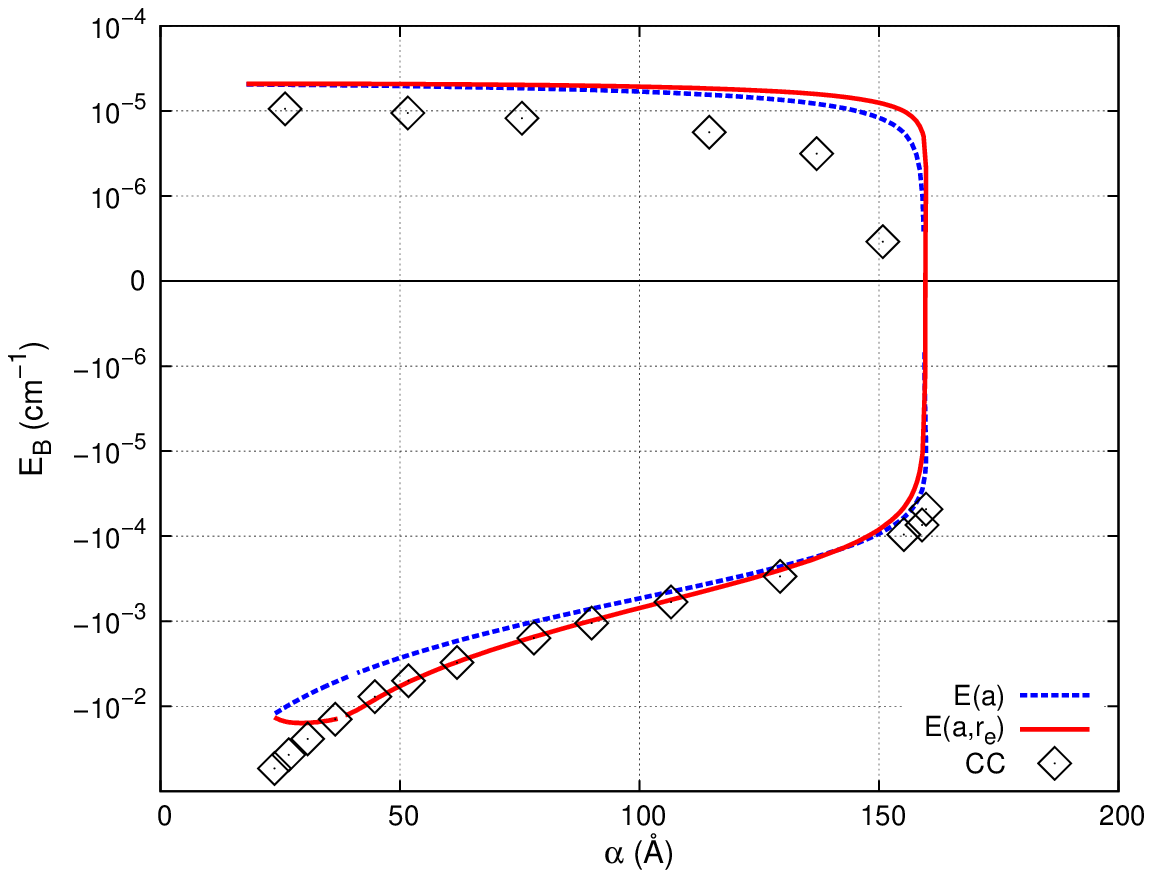}
\includegraphics[width=0.9\linewidth]{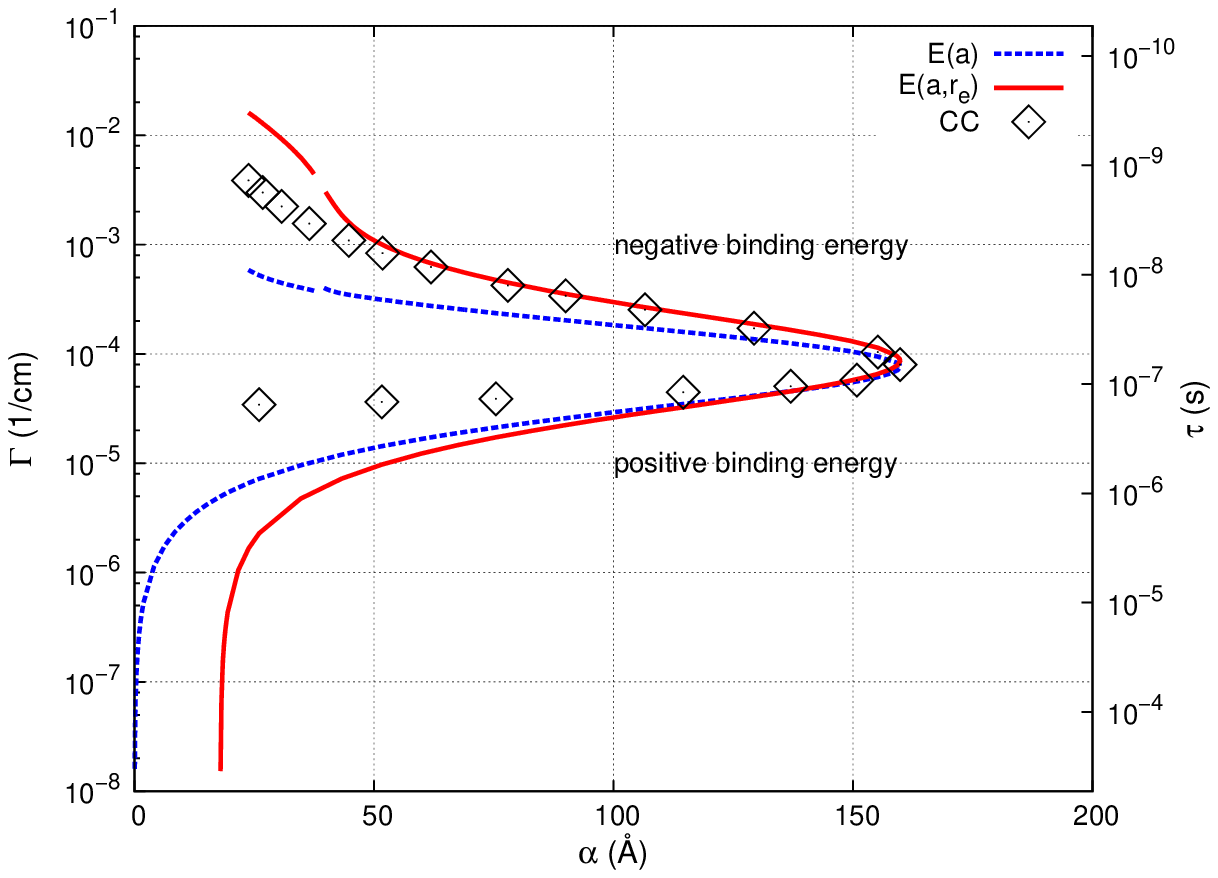}
\caption{The binding energy (upper panel) and the lifetime (lower panel) of the most weakly bound 
predissociating state as functions of the real part of the scattering length. The close coupling calculations are represented by diamonds, the results of equation (\ref{eqn:a}) -- by the blue, dashed line, and the 
the results of equation (\ref{eqn:aandr}) -- by the red, solid line.}
\label{fig:lifetimes}
\end{center}
\end{figure}

The binding energy, the width and the lifetime of the most weakly bound state are shown in Figure~\ref{fig:lifetimes} as functions of the real part of the scattering length. 
It can be seen from the lower panel of Figure~\ref{fig:lifetimes} that the calculation based on equation (\ref{eqn:aandr}) agrees with the close coupling numerical calculations far better than the calculation based on equation (\ref{eqn:a})
for $\alpha \gtrsim 50$~\AA~when the binding energy is negative. 
As the bound state moves above threshold the extrapolation of $E_{\rm B}$ and $\Gamma$ from the eigenphase sum below threshold breaks down 
and the calculated results become inaccurate.
The gap in the data is due to the presence of the additional $n=2$ Feshbach resonance at  $\lambda_\text{scale} \approx1.065$.

Starting from the bound state existing below threshold, as we decrease $\lambda_\text{scale}$, increasing $\alpha$, the binding energy decreases and the predissociation lifetime increases. 
When $\alpha\approx\beta\approx160$~\AA, the bound state crosses threshold, which is represented by the sharp increase of the curves in the upper panel of Figure~\ref{fig:lifetimes}. Decreasing $\lambda_\text{scale}$ further, decreases $\alpha$ and moves the bound state above threshold. At low scattering lengths, the width of the state shows a dramatic decrease before becoming negative, at which point 
the bound state becomes virtual.  At these small positive scattering lengths, it can be seen that the Mg-NH bound state lifetime can be on the order of tens to hundreds of microseconds. From Eq (7) $\Gamma\rightarrow0$ as $\alpha\rightarrow 0^+$, however with the effective range correction Eq (14), $\Gamma\rightarrow0$ when $\alpha\approx18$ \AA. Thus in principle the width of the state can be arbitrarily small. 

\begin{figure}[t]
\begin{center}
\includegraphics[width=0.5\linewidth]{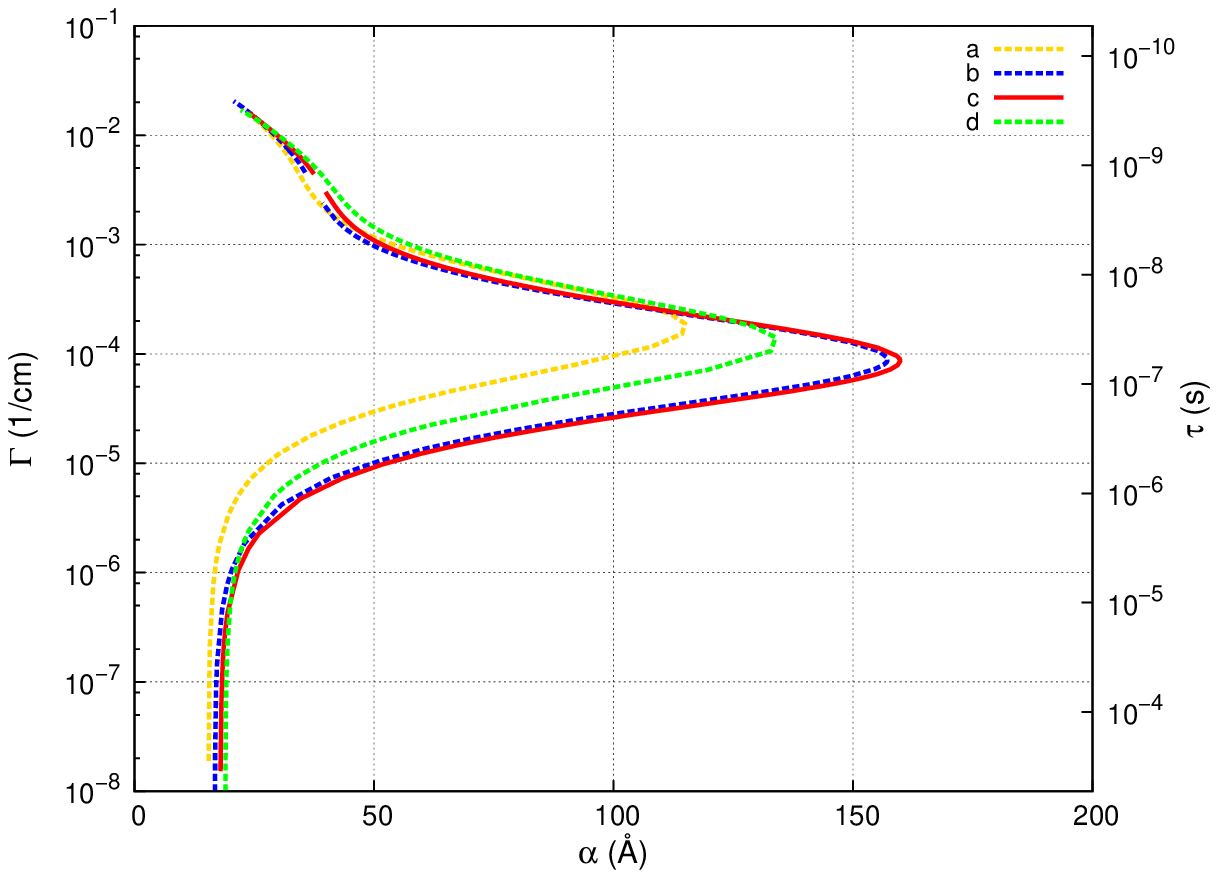}
\includegraphics[width=0.5\linewidth]{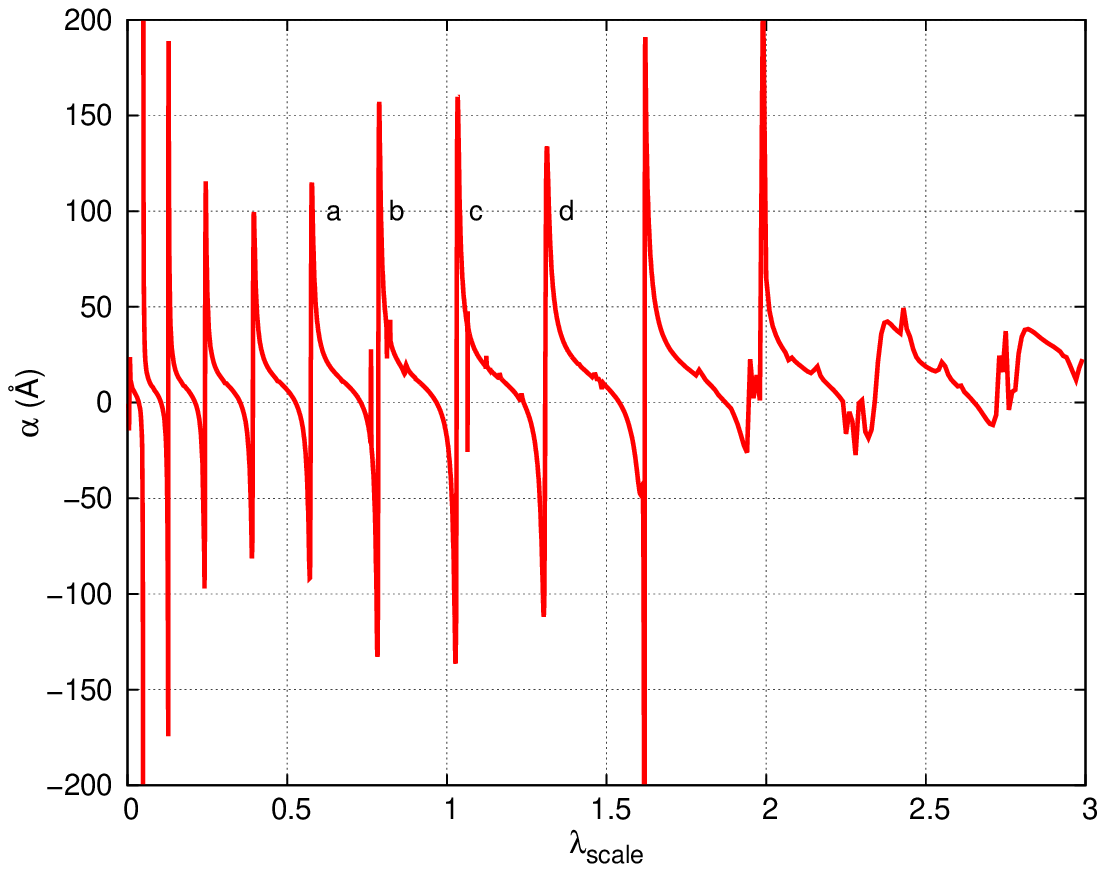}
\includegraphics[width=0.5\linewidth]{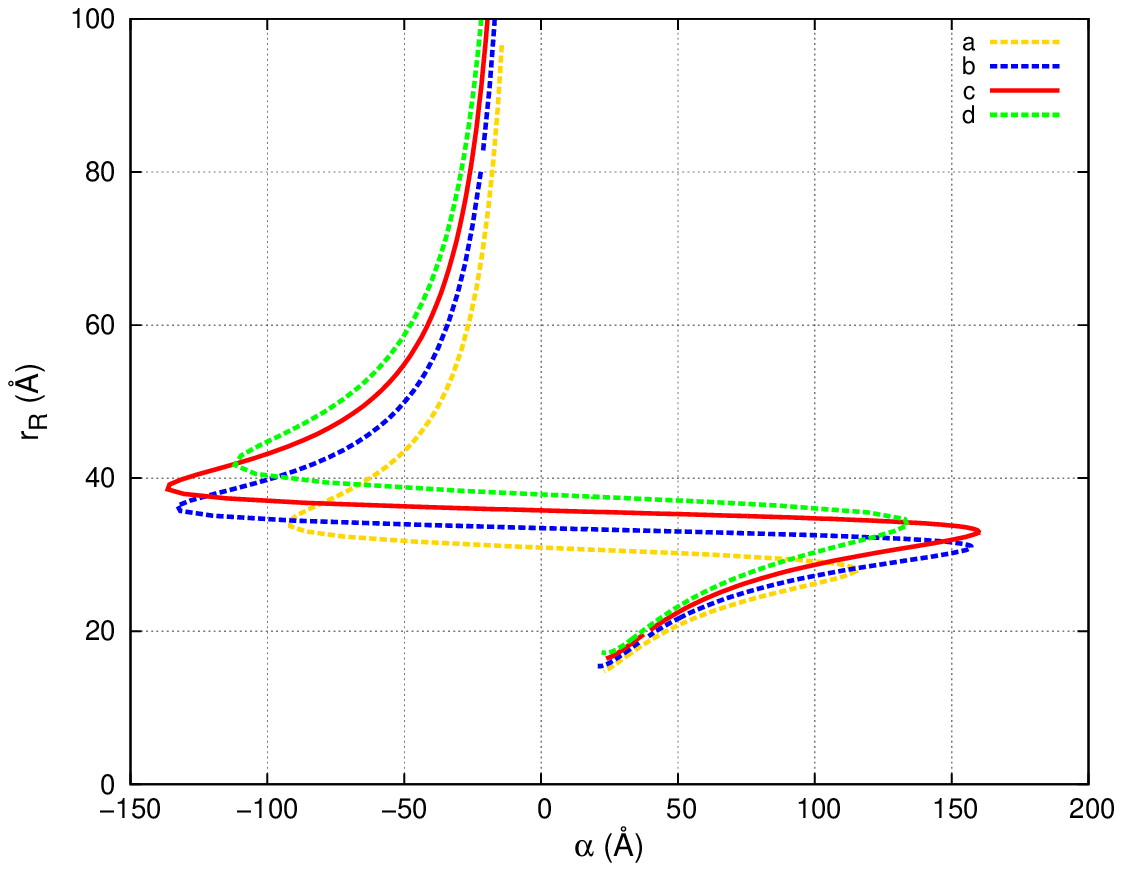}
\caption{Upper panel: The lifetime of the near-threshold predissociating state as a function of the scattering length modified by varying the interactional potential using different ranges of the scaling parameters corresponding to different peaks in the middle panel. Middle panel: The real part of 
the Mg+NH($n=1,j=1$) $s$-wave scattering length as a function of the interaction potential scaling parameter. Lower panel: The real part of the effective range for the Mg+NH($n=1,j=1$) collision complex as a function of the scattering length corresponding to different peaks in the middle panel. 
}
\label{fig:longvsshort}
\end{center}
\end{figure}

The results presented in Figure 2 are specific for the Mg - NH complex. However, Eqs. (13) and (14) are universal. To explore the universality of the predissoication lifetimes presented in Figure 2, we repeated the calculations for different values of the interaction potential scaling parameter $\lambda_{\rm scale}$. As $\lambda_{\rm scale}$ increases, the strength of the interaction potential increases. This increases the number of bound energy levels supported by the potential. The scattering length as a function of $\lambda_{\rm scale}$ exhibits a series of peaks corresponding to zero-energy bound states (see the middle panel of Figure 3). The upper panel of Figure 3 displays the predissociation lifetimes as functions of the scattering lengths calculated using different ranges of the scaling parameter corresponding to different peaks in the middle panel of Figure 3. The different curves in the upper panel of Figure 3 thus correspond to different interaction strengths. It can be seen that that the predissociation lifetimes of the bound states very near threshold are sensitive to the interaction strength. This is  because the magnitude of the effective range is different for different interaction strengths, as demonstrated in the lower panel of Figure 3. However,  the predissociation lifetimes remain similar at small values of the scattering length. In particular, the large values of the predissociation lifetimes corresponding to the bound state above threshold appear to be insensitive to the interaction strength. This indicates that the large magnitude of the predissociation lifetimes just above the excited state threshold is a general phenomenon.

\begin{figure}[t]
\begin{center}
\includegraphics[width=0.95\linewidth]{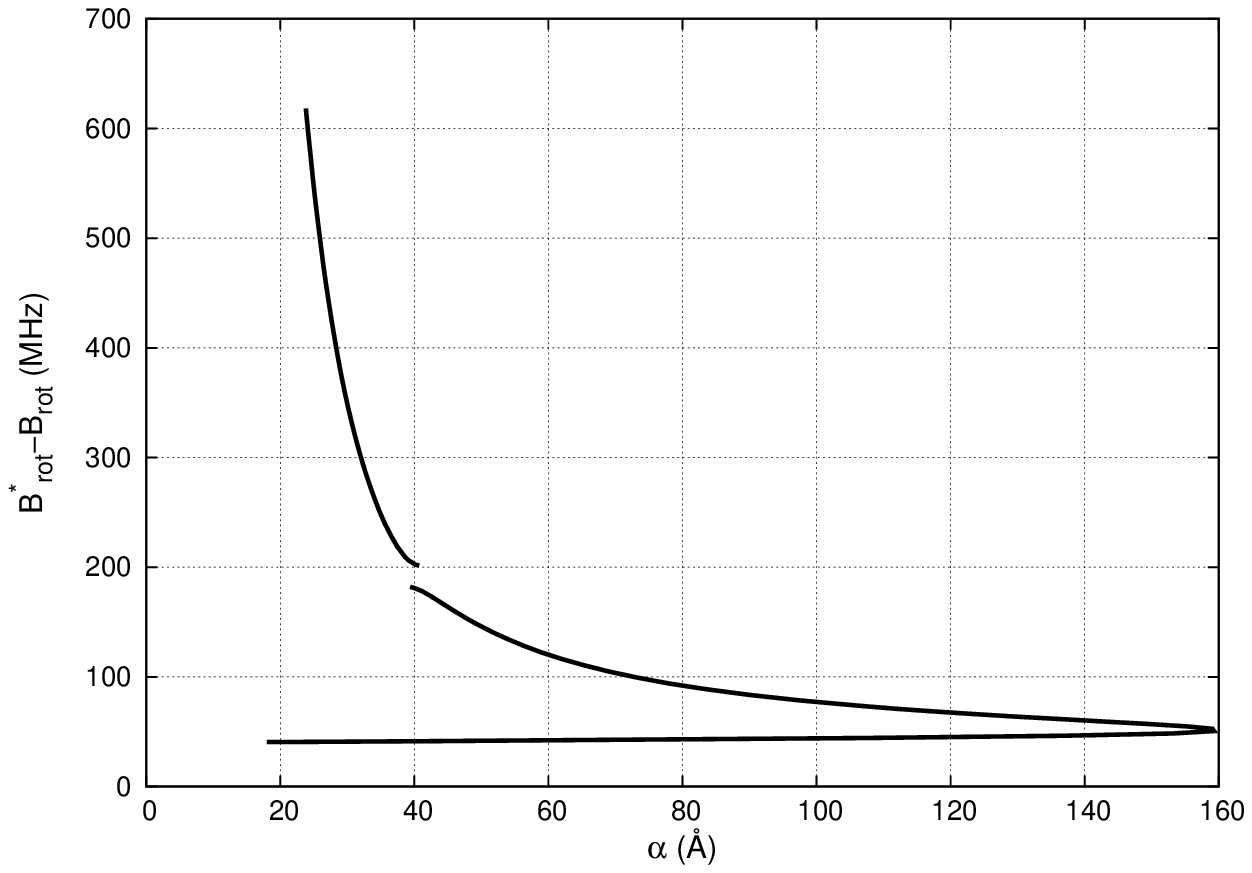}
\caption{The perturbation of the NH rotational constant due the presence of the 
weakly bound Mg atom, shown as a function of the real part of the scattering length.}
\label{fig:effB}
\end{center}
\end{figure}


The presence of an atom weakly bound to a molecule perturbs the rotational energy structure of the molecule. Atom - molecule Feshbach resonances can be used to introduce controllable perturbations to the rotational structure of ultracold molecules. This could be used to create an ordered array of ultracold molecules on an optical lattice  in the presence of  tunable disorder potential. It is therefore important to quantify the effect of the atom - molecule potential on the rotational structure of the molecules in Feshbach-resonance associated atom - molecule complexes. 

The energy separation of the NH($n=0,j=1$) rotational ground state and NH($n=1,j=1$) can be approximated as
\begin{equation}
(E_\text{n1j1}-E_\text{n0j1})\approx2B_\text{rot}n(n+1)-\gamma_\text{sn}+\frac{2\lambda_\text{ss}}{3},
\end{equation}
where in this approximation the rotational constant is $B_\text{rot}=16.34685$ cm$^{-1}$ rather than the spectroscopic value of 16.343 cm$^{-1}$ and the spin-rotation $\gamma_\text{sn}$ and spin-spin $\lambda_\text{ss}$ constants are -0.055 and 0.920 cm$^{-1}$ respectively \cite{Mizushima}. When the Mg atom is present the energy separation is modified, resulting in an effective rotational constant  $B_\text{rot}^*$. Figure~\ref{fig:effB} shows the energy difference between $B_\text{rot}^*$ and $B_\text{rot}$ calculated using Eq. (\ref{eqn:aandr}), as a function of $\alpha$. It can be seen that the rotational constant of the molecule can be effectively modified by linking the atom and the molecule with a Feshbach resonance.  




\section{Summary}

We have shown that multichannel effective range theory can be used to calculate the location and lifetime of the most weakly bound predissociating state near threshold, even in the presence of strong couplings between molecular states (strong inelasticity).  As the predissociating state is tuned closer to and moves across threshold, the lifetime of the state increases. We have found that the predissociation lifetime of the states just {\it above} the excited state threshold is the largest and can be tens to hundreds of microseconds long. For illustrative purposes, the location of the bound state was tuned in the present work by varying the depth of the potential. In experiments, bound states can be tuned across threshold using externally applied electric and magnetic fields \cite{Krems:book:2009}.
%
%

An ensemble of molecules on an optical lattice can be used to study collective excitations characteristic of solid-state crystals \cite{Herrera:2010,  Perez:2010,Herrera:excitionphonon:2010}.  The presence of an atom weakly bound to a molecule creates a perturbation in the rotational energy structure of the molecule. Our calculation shows that this perturbation can be tuned from zero to  $\sim 100$ MHz. 
Adding ultracold atoms to certain lattice sites and using Feshbach resonances could thus be exploited to modify the rotational structure of selected molecules trapped on an optical lattice. Rotational excitation of molecules in such a system would resemble collective excitations in a crystal with impurities \cite{Herrera:2010}. The characteristic time for transfer of rotational excitation energy between molecules in different sites of an optical lattice is $\sim 10~ \mu$s. The calculations of the present work show that the lifetimes for the rotational predissociation can be as large as $\sim 1$ ms.

 Mixtures of ultracold atoms and molecules have recently been created in several experimental laboratories \cite{Ospelkaus:science:2010,Knoop:atomdimer:2010}. The ultracold atom - molecule complexes undergoing the rotational predissociation studied in this work can be created by first linking ultracold molecules in the ground rotational state with ultracold atoms using a Feshbach resonance and then exciting the molecules to a rotationally excited state. The external field used for the Feshbach resonance association can be varied to adjust the binding energy of the atom - molecule complex. We hope that the results of the present work will stimulate the measurements of near-threshold rotational predissociation.

\section{Acknowledgements}
The work was supported by NSERC of Canada.


\begin{thebibliography}{39}
\expandafter\ifx\csname natexlab\endcsname\relax\def\natexlab#1{#1}\fi
\expandafter\ifx\csname bibnamefont\endcsname\relax
  \def\bibnamefont#1{#1}\fi
\expandafter\ifx\csname bibfnamefont\endcsname\relax
  \def\bibfnamefont#1{#1}\fi
\expandafter\ifx\csname citenamefont\endcsname\relax
  \def\citenamefont#1{#1}\fi
\expandafter\ifx\csname url\endcsname\relax
  \def\url#1{\texttt{#1}}\fi
\expandafter\ifx\csname urlprefix\endcsname\relax\def\urlprefix{URL }\fi
\providecommand{\bibinfo}[2]{#2}
\providecommand{\eprint}[2][]{\url{#2}}

\bibitem[{\citenamefont{Doyle et~al.}(2004)\citenamefont{Doyle, Friedrich,
  Krems, and Masnou-Seeuws}}]{Doyle:2004}
\bibinfo{author}{\bibfnamefont{J.}~\bibnamefont{Doyle}},
  \bibinfo{author}{\bibfnamefont{B.}~\bibnamefont{Friedrich}},
  \bibinfo{author}{\bibfnamefont{R.~V.} \bibnamefont{Krems}}, \bibnamefont{and}
  \bibinfo{author}{\bibfnamefont{F.}~\bibnamefont{Masnou-Seeuws}},
  \bibinfo{journal}{Eur. Phys. J. D} \textbf{\bibinfo{volume}{31}},
  \bibinfo{pages}{149} (\bibinfo{year}{2004}).

\bibitem[{\citenamefont{Hutson and Sold\'{a}n}(2006)}]{Hutson:IRPC:2006}
\bibinfo{author}{\bibfnamefont{J.~M.} \bibnamefont{Hutson}} \bibnamefont{and}
  \bibinfo{author}{\bibfnamefont{P.}~\bibnamefont{Sold\'{a}n}},
  \bibinfo{journal}{Int. Rev. Phys. Chem.} \textbf{\bibinfo{volume}{25}},
  \bibinfo{pages}{497} (\bibinfo{year}{2006}).

\bibitem[{\citenamefont{K\"{o}hler et~al.}(2006)\citenamefont{K\"{o}hler,
  Goral, and Julienne}}]{Kohler:RMP:2006}
\bibinfo{author}{\bibfnamefont{T.}~\bibnamefont{K\"{o}hler}},
  \bibinfo{author}{\bibfnamefont{K.}~\bibnamefont{Goral}}, \bibnamefont{and}
  \bibinfo{author}{\bibfnamefont{P.~S.} \bibnamefont{Julienne}},
  \bibinfo{journal}{Rev. Mod. Phys.} \textbf{\bibinfo{volume}{78}},
  \bibinfo{pages}{1311} (\bibinfo{year}{2006}).

\bibitem[{\citenamefont{Chin et~al.}(2010)\citenamefont{Chin, Grimm, Julienne,
  and Tiesinga}}]{Chin:RMP:2010}
\bibinfo{author}{\bibfnamefont{C.}~\bibnamefont{Chin}},
  \bibinfo{author}{\bibfnamefont{R.}~\bibnamefont{Grimm}},
  \bibinfo{author}{\bibfnamefont{P.}~\bibnamefont{Julienne}}, \bibnamefont{and}
  \bibinfo{author}{\bibfnamefont{E.}~\bibnamefont{Tiesinga}},
  \bibinfo{journal}{Rev. Mod. Phys.} \textbf{\bibinfo{volume}{82}},
  \bibinfo{pages}{1225} (\bibinfo{year}{2010}).

\bibitem[{\citenamefont{Carr et~al.}(2009)\citenamefont{Carr, {DeMille}, Krems,
  and Ye}}]{Carr:NJPintro:2009}
\bibinfo{author}{\bibfnamefont{L.~D.} \bibnamefont{Carr}},
  \bibinfo{author}{\bibfnamefont{D.}~\bibnamefont{{DeMille}}},
  \bibinfo{author}{\bibfnamefont{R.~V.} \bibnamefont{Krems}}, \bibnamefont{and}
  \bibinfo{author}{\bibfnamefont{J.}~\bibnamefont{Ye}}, \bibinfo{journal}{New
  J. Phys.} \textbf{\bibinfo{volume}{11}}, \bibinfo{pages}{055049}
  (\bibinfo{year}{2009}).

\bibitem[{\citenamefont{Chin et~al.}(2005)\citenamefont{Chin, Kraemer, Mark,
  Herbig, Waldburger, N\"{a}gerl, and Grimm}}]{Chin:2005}
\bibinfo{author}{\bibfnamefont{C.}~\bibnamefont{Chin}},
  \bibinfo{author}{\bibfnamefont{T.}~\bibnamefont{Kraemer}},
  \bibinfo{author}{\bibfnamefont{M.}~\bibnamefont{Mark}},
  \bibinfo{author}{\bibfnamefont{J.}~\bibnamefont{Herbig}},
  \bibinfo{author}{\bibfnamefont{P.}~\bibnamefont{Waldburger}},
  \bibinfo{author}{\bibfnamefont{H.~C.} \bibnamefont{N\"{a}gerl}},
  \bibnamefont{and} \bibinfo{author}{\bibfnamefont{R.}~\bibnamefont{Grimm}},
  \bibinfo{journal}{Phys. Rev. Lett.} \textbf{\bibinfo{volume}{94}},
  \bibinfo{pages}{123201} (\bibinfo{year}{2005}).

\bibitem[{\citenamefont{Beswick and Jortner}(1978)}]{Beswick:1978}
\bibinfo{author}{\bibfnamefont{J.~A.} \bibnamefont{Beswick}} \bibnamefont{and}
  \bibinfo{author}{\bibfnamefont{J.}~\bibnamefont{Jortner}},
  \bibinfo{journal}{J. Chem. Phys.} \textbf{\bibinfo{volume}{68}},
  \bibinfo{pages}{2277} (\bibinfo{year}{1978}).

\bibitem[{\citenamefont{Pine and Lafferty}(1983)}]{Pine:1983}
\bibinfo{author}{\bibfnamefont{A.}~\bibnamefont{Pine}} \bibnamefont{and}
  \bibinfo{author}{\bibfnamefont{W.}~\bibnamefont{Lafferty}},
  \bibinfo{journal}{J. Chem. Phys.} \textbf{\bibinfo{volume}{78}},
  \bibinfo{pages}{2154} (\bibinfo{year}{1983}).

\bibitem[{\citenamefont{Ashton et~al.}(1983)\citenamefont{Ashton, Child, and
  Hutson}}]{Ashton:1983}
\bibinfo{author}{\bibfnamefont{C.~J.} \bibnamefont{Ashton}},
  \bibinfo{author}{\bibfnamefont{M.~S.} \bibnamefont{Child}}, \bibnamefont{and}
  \bibinfo{author}{\bibfnamefont{J.~M.} \bibnamefont{Hutson}},
  \bibinfo{journal}{J. Chem. Phys.} \textbf{\bibinfo{volume}{78}},
  \bibinfo{pages}{4025} (\bibinfo{year}{1983}).

\bibitem[{\citenamefont{Giancarlo et~al.}(1994)\citenamefont{Giancarlo,
  Randall, Choi, and Lester}}]{giancarlo:1994}
\bibinfo{author}{\bibfnamefont{L.}~\bibnamefont{Giancarlo}},
  \bibinfo{author}{\bibfnamefont{R.}~\bibnamefont{Randall}},
  \bibinfo{author}{\bibfnamefont{S.}~\bibnamefont{Choi}}, \bibnamefont{and}
  \bibinfo{author}{\bibfnamefont{M.}~\bibnamefont{Lester}},
  \bibinfo{journal}{J. Chem. Phys.} \textbf{\bibinfo{volume}{101}},
  \bibinfo{pages}{2914} (\bibinfo{year}{1994}).

\bibitem[{\citenamefont{Lang et~al.}(2008)\citenamefont{Lang, Winkler, Strauss,
  Grimm, and Hecker~Denschlag}}]{Lang:ground:2008}
\bibinfo{author}{\bibfnamefont{F.}~\bibnamefont{Lang}},
  \bibinfo{author}{\bibfnamefont{K.}~\bibnamefont{Winkler}},
  \bibinfo{author}{\bibfnamefont{C.}~\bibnamefont{Strauss}},
  \bibinfo{author}{\bibfnamefont{R.}~\bibnamefont{Grimm}}, \bibnamefont{and}
  \bibinfo{author}{\bibfnamefont{J.}~\bibnamefont{Hecker~Denschlag}},
  \bibinfo{journal}{Phys. Rev. Lett.} \textbf{\bibinfo{volume}{101}},
  \bibinfo{pages}{133005} (\bibinfo{year}{2008}).

\bibitem[{\citenamefont{Ni et~al.}(2008)\citenamefont{Ni, Ospelkaus, {de
  Miranda}, Pe'er, Neyenhuis, Zirbel, Kotochigova, Julienne, Jin, and
  Ye}}]{Ni:KRb:2008}
\bibinfo{author}{\bibfnamefont{K.-K.} \bibnamefont{Ni}},
  \bibinfo{author}{\bibfnamefont{S.}~\bibnamefont{Ospelkaus}},
  \bibinfo{author}{\bibfnamefont{M.~H.~G.} \bibnamefont{{de Miranda}}},
  \bibinfo{author}{\bibfnamefont{A.}~\bibnamefont{Pe'er}},
  \bibinfo{author}{\bibfnamefont{B.}~\bibnamefont{Neyenhuis}},
  \bibinfo{author}{\bibfnamefont{J.~J.} \bibnamefont{Zirbel}},
  \bibinfo{author}{\bibfnamefont{S.}~\bibnamefont{Kotochigova}},
  \bibinfo{author}{\bibfnamefont{P.~S.} \bibnamefont{Julienne}},
  \bibinfo{author}{\bibfnamefont{D.~S.} \bibnamefont{Jin}}, \bibnamefont{and}
  \bibinfo{author}{\bibfnamefont{J.}~\bibnamefont{Ye}},
  \bibinfo{journal}{Science} \textbf{\bibinfo{volume}{322}},
  \bibinfo{pages}{231} (\bibinfo{year}{2008}).

\bibitem[{\citenamefont{Herrera et~al.}(2010)\citenamefont{Herrera, Litinskaya,
  and Krems}}]{Herrera:2010}
\bibinfo{author}{\bibfnamefont{F.}~\bibnamefont{Herrera}},
  \bibinfo{author}{\bibfnamefont{M.}~\bibnamefont{Litinskaya}},
  \bibnamefont{and} \bibinfo{author}{\bibfnamefont{R.~V.} \bibnamefont{Krems}},
  \bibinfo{journal}{Phys. Rev. A} \textbf{\bibinfo{volume}{82}},
  \bibinfo{pages}{033428} (\bibinfo{year}{2010}).

\bibitem[{\citenamefont{P{\'e}rez-R{\'\i}os
  et~al.}(2010)\citenamefont{P{\'e}rez-R{\'\i}os, Herrera, and
  Krems}}]{Perez:2010}
\bibinfo{author}{\bibfnamefont{J.}~\bibnamefont{P{\'e}rez-R{\'\i}os}},
  \bibinfo{author}{\bibfnamefont{F.}~\bibnamefont{Herrera}}, \bibnamefont{and}
  \bibinfo{author}{\bibfnamefont{R.~V.} \bibnamefont{Krems}},
  \bibinfo{journal}{New J. Phys.} \textbf{\bibinfo{volume}{12}},
  \bibinfo{pages}{103007} (\bibinfo{year}{2010}).

\bibitem[{\citenamefont{Herrera and Krems}(2010)}]{Herrera:excitionphonon:2010}
\bibinfo{author}{\bibfnamefont{F.}~\bibnamefont{Herrera}} \bibnamefont{and}
  \bibinfo{author}{\bibfnamefont{R.}~\bibnamefont{Krems}},
  \bibinfo{journal}{arXiv:physics/1010.1782}  (\bibinfo{year}{2010}).

\bibitem[{\citenamefont{{Le Roy} et~al.}(1982)\citenamefont{{Le Roy}, Corey,
  and Hutson}}]{LeRoy:1982}
\bibinfo{author}{\bibfnamefont{R.~J.} \bibnamefont{{Le Roy}}},
  \bibinfo{author}{\bibfnamefont{G.~C.} \bibnamefont{Corey}}, \bibnamefont{and}
  \bibinfo{author}{\bibfnamefont{J.~M.} \bibnamefont{Hutson}},
  \bibinfo{journal}{Faraday Discuss. Chem. Soc.} \textbf{\bibinfo{volume}{73}},
  \bibinfo{pages}{339} (\bibinfo{year}{1982}).

\bibitem[{\citenamefont{Hutson et~al.}(1983)\citenamefont{Hutson, Ashton, and
  {Le Roy}}}]{HUTSON:ArH2:1983}
\bibinfo{author}{\bibfnamefont{J.~M.} \bibnamefont{Hutson}},
  \bibinfo{author}{\bibfnamefont{C.~J.} \bibnamefont{Ashton}},
  \bibnamefont{and} \bibinfo{author}{\bibfnamefont{R.~J.} \bibnamefont{{Le
  Roy}}}, \bibinfo{journal}{J. Phys. Chem.} \textbf{\bibinfo{volume}{87}},
  \bibinfo{pages}{2713} (\bibinfo{year}{1983}).

\bibitem[{\citenamefont{Forrey et~al.}(1999)\citenamefont{Forrey, Kharchenko,
  Balakrishnan, and Dalgarno}}]{Forrey:vib:1999}
\bibinfo{author}{\bibfnamefont{R.~C.} \bibnamefont{Forrey}},
  \bibinfo{author}{\bibfnamefont{V.}~\bibnamefont{Kharchenko}},
  \bibinfo{author}{\bibfnamefont{N.}~\bibnamefont{Balakrishnan}},
  \bibnamefont{and} \bibinfo{author}{\bibfnamefont{A.}~\bibnamefont{Dalgarno}},
  \bibinfo{journal}{Phys. Rev. A} \textbf{\bibinfo{volume}{59}},
  \bibinfo{pages}{2146} (\bibinfo{year}{1999}).

\bibitem[{\citenamefont{Balakrishnan et~al.}(1997)\citenamefont{Balakrishnan,
  Kharchenko, Forrey, and Dalgarno}}]{Balakrishnan:scat-len:1997}
\bibinfo{author}{\bibfnamefont{N.}~\bibnamefont{Balakrishnan}},
  \bibinfo{author}{\bibfnamefont{V.}~\bibnamefont{Kharchenko}},
  \bibinfo{author}{\bibfnamefont{R.~C.} \bibnamefont{Forrey}},
  \bibnamefont{and} \bibinfo{author}{\bibfnamefont{A.}~\bibnamefont{Dalgarno}},
  \bibinfo{journal}{Chem. Phys. Lett.} \textbf{\bibinfo{volume}{280}},
  \bibinfo{pages}{5} (\bibinfo{year}{1997}).

\bibitem[{\citenamefont{Forrey et~al.}(1998)\citenamefont{Forrey, Balakrishnan,
  Kharchenko, and Dalgarno}}]{Forrey:1998}
\bibinfo{author}{\bibfnamefont{R.~C.} \bibnamefont{Forrey}},
  \bibinfo{author}{\bibfnamefont{N.}~\bibnamefont{Balakrishnan}},
  \bibinfo{author}{\bibfnamefont{V.}~\bibnamefont{Kharchenko}},
  \bibnamefont{and} \bibinfo{author}{\bibfnamefont{A.}~\bibnamefont{Dalgarno}},
  \bibinfo{journal}{Phys. Rev. A} \textbf{\bibinfo{volume}{58}},
  \bibinfo{pages}{R2645} (\bibinfo{year}{1998}).

\bibitem[{\citenamefont{Mott and Massey}(1965)}]{Mott:1965}
\bibinfo{author}{\bibfnamefont{N.~F.} \bibnamefont{Mott}} \bibnamefont{and}
  \bibinfo{author}{\bibfnamefont{H.~S.~W.} \bibnamefont{Massey}},
  \emph{\bibinfo{title}{The Theory of Atomic Collisions}}
  (\bibinfo{publisher}{Clarendon Press}, \bibinfo{address}{Oxford},
  \bibinfo{year}{1965}), \bibinfo{edition}{3rd} ed.

\bibitem[{\citenamefont{Ross and Shaw}(1960)}]{Ross:1960}
\bibinfo{author}{\bibfnamefont{M.~H.} \bibnamefont{Ross}} \bibnamefont{and}
  \bibinfo{author}{\bibfnamefont{G.~L.} \bibnamefont{Shaw}},
  \bibinfo{journal}{Annals of Physics} \textbf{\bibinfo{volume}{9}},
  \bibinfo{pages}{391} (\bibinfo{year}{1960}).

\bibitem[{\citenamefont{Ross and Shaw}(1961)}]{Ross:1961}
\bibinfo{author}{\bibfnamefont{M.~H.} \bibnamefont{Ross}} \bibnamefont{and}
  \bibinfo{author}{\bibfnamefont{G.~L.} \bibnamefont{Shaw}},
  \bibinfo{journal}{Annals of Physics} \textbf{\bibinfo{volume}{13}},
  \bibinfo{pages}{147} (\bibinfo{year}{1961}).

\bibitem[{\citenamefont{Bohn and Julienne}(1997)}]{Bohn:1997}
\bibinfo{author}{\bibfnamefont{J.~L.} \bibnamefont{Bohn}} \bibnamefont{and}
  \bibinfo{author}{\bibfnamefont{P.~S.} \bibnamefont{Julienne}},
  \bibinfo{journal}{Phys. Rev. A} \textbf{\bibinfo{volume}{56}},
  \bibinfo{pages}{1486} (\bibinfo{year}{1997}).

\bibitem[{\citenamefont{Hutson}(2007{\natexlab{a}})}]{Hutson:res:nonote:2007}
\bibinfo{author}{\bibfnamefont{J.~M.} \bibnamefont{Hutson}},
  \bibinfo{journal}{New J. Phys.} \textbf{\bibinfo{volume}{9}},
  \bibinfo{pages}{152} (\bibinfo{year}{2007}{\natexlab{a}}).

\bibitem[{\citenamefont{Taylor}(1972)}]{Taylor:1972}
\bibinfo{author}{\bibfnamefont{J.~R.} \bibnamefont{Taylor}},
  \emph{\bibinfo{title}{Scattering Theory: The Quantum Theory of
  Nonrelativistic Collisions}} (\bibinfo{publisher}{Wiley},
  \bibinfo{address}{New York}, \bibinfo{year}{1972}).

\bibitem[{\citenamefont{Gonz\'{a}lez-Mart\'{\i}nez and
  Hutson}(2007)}]{Gonzalez-Martinez:2007}
\bibinfo{author}{\bibfnamefont{M.~L.} \bibnamefont{Gonz\'{a}lez-Mart\'{\i}nez}}
  \bibnamefont{and} \bibinfo{author}{\bibfnamefont{J.~M.}
  \bibnamefont{Hutson}}, \bibinfo{journal}{Phys. Rev. A}
  \textbf{\bibinfo{volume}{75}}, \bibinfo{pages}{022702}
  (\bibinfo{year}{2007}).


\bibitem [{\citenamefont {Moerdijk}, \citenamefont {Verhaar},\ and\
  \citenamefont {Axelsson}(1995)}]{Moerdijk:1995}
  \bibinfo {author} {\bibfnamefont {A.~J.}~\bibnamefont
  {Moerdijk}}, \bibinfo {author} {\bibfnamefont {B.~J.}~\bibnamefont
  {Verhaar}}, \bibnamefont{and} \bibinfo {author} {\bibfnamefont {A.}~\bibnamefont
  {Axelsson}}, \bibinfo  {journal} {Phys. Rev. A} \textbf {\bibinfo {volume} {51}}, \bibinfo {pages} {4852}
  (\bibinfo {year} {1995}).
  
\bibitem[{\citenamefont{Sold\'{a}n et~al.}(2009)\citenamefont{Sold\'{a}n,
  \.Zuchowski, and Hutson}}]{Soldan:MgNH:2009}
\bibinfo{author}{\bibfnamefont{P.}~\bibnamefont{Sold\'{a}n}},
  \bibinfo{author}{\bibfnamefont{P.~S.} \bibnamefont{\.Zuchowski}},
  \bibnamefont{and} \bibinfo{author}{\bibfnamefont{J.~M.}
  \bibnamefont{Hutson}}, \bibinfo{journal}{Faraday Discuss.}
  \textbf{\bibinfo{volume}{142}}, \bibinfo{pages}{191} (\bibinfo{year}{2009}).

\bibitem[{\citenamefont{Wallis and Hutson}(2009)}]{Wallis:PRL:MgNH:2009}
\bibinfo{author}{\bibfnamefont{A.~O.~G.} \bibnamefont{Wallis}}
  \bibnamefont{and} \bibinfo{author}{\bibfnamefont{J.~M.}
  \bibnamefont{Hutson}}, \bibinfo{journal}{Phys. Rev. Lett.}
  \textbf{\bibinfo{volume}{103}}, \bibinfo{pages}{183201}
  (\bibinfo{year}{2009}).

\bibitem[{\citenamefont{Wallis}(2010)}]{Wallis:phd:2010}
\bibinfo{author}{\bibfnamefont{A.~O.~G.} \bibnamefont{Wallis}},
  \bibinfo{journal}{PhD Thesis, Durham University}  (\bibinfo{year}{2010}),
  \urlprefix\url{http://etheses.dur.ac.uk/184/}.

\bibitem[{\citenamefont{Hutson and Green}(1994)}]{molscat:v14}
\bibinfo{author}{\bibfnamefont{J.~M.} \bibnamefont{Hutson}} \bibnamefont{and}
  \bibinfo{author}{\bibfnamefont{S.}~\bibnamefont{Green}},
  \emph{\bibinfo{title}{{MOLSCAT} computer program, version 14}},
  \bibinfo{howpublished}{distributed by Collaborative Computational Project
  No.\ 6 of the UK Engineering and Physical Sciences Research Council}
  (\bibinfo{year}{1994}).

\bibitem[{\citenamefont{Alexander and Manolopoulos}(1987)}]{Alexander:1987}
\bibinfo{author}{\bibfnamefont{M.~H.} \bibnamefont{Alexander}}
  \bibnamefont{and} \bibinfo{author}{\bibfnamefont{D.~E.}
  \bibnamefont{Manolopoulos}}, \bibinfo{journal}{J. Chem. Phys.}
  \textbf{\bibinfo{volume}{86}}, \bibinfo{pages}{2044} (\bibinfo{year}{1987}).

\bibitem[{\citenamefont{Johnson}(1973)}]{Johnson:1973}
\bibinfo{author}{\bibfnamefont{B.~R.} \bibnamefont{Johnson}},
  \bibinfo{journal}{J. Comput. Phys.} \textbf{\bibinfo{volume}{13}},
  \bibinfo{pages}{445} (\bibinfo{year}{1973}).

\bibitem[{\citenamefont{Hutson}(2007{\natexlab{b}})}]{Hutson:resfit:2007}
\bibinfo{author}{\bibfnamefont{J.~M.} \bibnamefont{Hutson}},
  \emph{\bibinfo{title}{Resfit 2007 computer program}}
  (\bibinfo{year}{2007}{\natexlab{b}}).

\bibitem[{\citenamefont{Mizushima}(1975)}]{Mizushima}
\bibinfo{author}{\bibfnamefont{M.}~\bibnamefont{Mizushima}},
  \emph{\bibinfo{title}{Theory of Rotating Diatomic Molecules}}
  (\bibinfo{publisher}{Wiley}, \bibinfo{address}{New York},
  \bibinfo{year}{1975}).

\bibitem[{\citenamefont{Krems et~al.}(2009)\citenamefont{Krems, Stwalley, and
  Friedrich}}]{Krems:book:2009}
\bibinfo{author}{\bibfnamefont{R.~V.} \bibnamefont{Krems}},
  \bibinfo{author}{\bibfnamefont{W.~C.} \bibnamefont{Stwalley}},
  \bibnamefont{and}
  \bibinfo{author}{\bibfnamefont{B.}~\bibnamefont{Friedrich}},
  \emph{\bibinfo{title}{Cold Molecules: Theory, Experiment, Applications}}
  (\bibinfo{publisher}{CRC Press}, \bibinfo{address}{Boca Raton},
  \bibinfo{year}{2009}).

\bibitem[{\citenamefont{Ospelkaus et~al.}(2010)\citenamefont{Ospelkaus, Ni,
  Wang, De~Miranda, Neyenhuis, Qu{\'e}m{\'e}ner, Julienne, Bohn, Jin, and
  Ye}}]{Ospelkaus:science:2010}
\bibinfo{author}{\bibfnamefont{S.}~\bibnamefont{Ospelkaus}},
  \bibinfo{author}{\bibfnamefont{K.}~\bibnamefont{Ni}},
  \bibinfo{author}{\bibfnamefont{D.}~\bibnamefont{Wang}},
  \bibinfo{author}{\bibfnamefont{M.}~\bibnamefont{De~Miranda}},
  \bibinfo{author}{\bibfnamefont{B.}~\bibnamefont{Neyenhuis}},
  \bibinfo{author}{\bibfnamefont{G.}~\bibnamefont{Qu{\'e}m{\'e}ner}},
  \bibinfo{author}{\bibfnamefont{P.}~\bibnamefont{Julienne}},
  \bibinfo{author}{\bibfnamefont{J.}~\bibnamefont{Bohn}},
  \bibinfo{author}{\bibfnamefont{D.}~\bibnamefont{Jin}}, \bibnamefont{and}
  \bibinfo{author}{\bibfnamefont{J.}~\bibnamefont{Ye}},
  \bibinfo{journal}{Science} \textbf{\bibinfo{volume}{327}},
  \bibinfo{pages}{853} (\bibinfo{year}{2010}).

\bibitem[{\citenamefont{Knoop et~al.}(2010)\citenamefont{Knoop, Ferlaino,
  Berninger, Mark, N\"agerl, Grimm, D'Incao, and Esry}}]{Knoop:atomdimer:2010}
\bibinfo{author}{\bibfnamefont{S.}~\bibnamefont{Knoop}},
  \bibinfo{author}{\bibfnamefont{F.}~\bibnamefont{Ferlaino}},
  \bibinfo{author}{\bibfnamefont{M.}~\bibnamefont{Berninger}},
  \bibinfo{author}{\bibfnamefont{M.}~\bibnamefont{Mark}},
  \bibinfo{author}{\bibfnamefont{H.-C.} \bibnamefont{N\"agerl}},
  \bibinfo{author}{\bibfnamefont{R.}~\bibnamefont{Grimm}},
  \bibinfo{author}{\bibfnamefont{J.~P.} \bibnamefont{D'Incao}},
  \bibnamefont{and} \bibinfo{author}{\bibfnamefont{B.~D.} \bibnamefont{Esry}},
  \bibinfo{journal}{Phys. Rev. Lett.} \textbf{\bibinfo{volume}{104}},
  \bibinfo{pages}{053201} (\bibinfo{year}{2010}).

\end{thebibliography}

\end{document}